\documentclass[slac_one]{revtex4}
\usepackage{graphicx}
\usepackage{fancyhdr}
\def\CP {\ensuremath{C\!P}}

\def\Lbabar{\mbox{{\LARGE\sl B}\hspace{-0.15em}{\Large\sl A}\hspace{-0.07em}{\LARGE\sl B}\hspace{-0.15em}{\Large\sl A\hspace{-0.02em}R}}}
\def\lbabar{\mbox{{\large\sl B}\hspace{-0.4em} {\normalsize\sl A}\hspace{-0.03em}{\large\sl B}\hspace{-0.4em} {\normalsize\sl A\hspace{-0.02em}R}}}
\def\babar{\mbox{\sl B\hspace{-0.4em} {\small\sl A}\hspace{-0.37em}
    \sl B\hspace{-0.4em} {\small\sl A\hspace{-0.02em}R}}}
\def\sbabar{\mbox{\sl B\hspace{-0.4em} {\scriptsize\sl
      A}\hspace{-0.37em} \sl B\hspace{-0.4em} {\scriptsize\sl
      A\hspace{-0.02em}R}}}

\def\sss{\scriptscriptstyle}
\def\barpd{{\raise.35ex\hbox{${\sss (}$}}--{\raise.35ex\hbox{${\sss )}$}}}
\def\dbarp{\hbox{$D^{0}$\kern-1.3em\raise1.5ex\hbox{\barpd}}\:}
\def\dbarp{\hbox{$D^{0}$\kern-1.3em\raise1.5ex\hbox{\barpd}}\:}

\begin{document}

\title{Measurements of the CKM angle $\gamma$ at \Lbabar}

%

\author{G. Marchiori, on behalf of the \babar\ Collaboration}
\affiliation{Universit\`a di Pisa, Dipartimento di Fisica, and INFN, Pisa, I-56127, Italy\\E-mail: {\it giovanni.marchiori@pi.infn.it}}

\begin{abstract}
We report on our recent measurements of the Cabibbo-Kobayashi-Maskawa
\CP-violating phase $\gamma$ and of related \CP-asymmetries and
branching fraction ratios. The measurements have been performed on
samples of up to 465 million $B\overline B$ pairs collected by the \sbabar\
detector at the SLAC PEP-II asymmetric-energy $B$ factory in the years
1999-2007.
\end{abstract}

\maketitle

\thispagestyle{fancy}


\section{INTRODUCTION}
Within the Standard Model \CP\ violation arises from a single irreducible
complex phase in the Cabibbo-Kobayashi-Maskawa (CKM) quark-flavor-mixing
matrix $V$
and manifests itself as a non-zero area of the Unitarity Triangle
(UT). 
This triangle depicts, in the complex plane, the relation
$1{+}\frac{V_{ud}V_{ub}^*}{V_{cd}V_{cb}^*}{+}\frac{V_{td}V^*_{tb}}{V_{cd}V_{cb}^*}{=}0$,
that follows from the unitarity of $V$.
The \babar\ experiment has measured precisely, in $B\to (c\bar c)X_s$
decays, the angle $\beta\equiv
\arg - \frac{V_{cd}V_{cb}^*}{V_{td}V_{tb}^*}$ of the UT,
finding a value that differs significantly from 0 and
$\pi$. This clearly indicates that the area of the UT is
non-vanishing, thus establishing \CP-violation.
In order to confirm that the CKM mechanism is the correct explanation
for \CP-violation, we need to overconstrain the UT by measuring
precisely the other angles ($\alpha$ and $\gamma$) and the sides.

\section{MEASURING $\gamma$ WITH $B$ MESON DECAYS IN \lbabar}
The angle $\gamma\equiv\arg -\frac{V_{ud}V_{ub}^*}{V_{cd}V_{cb}^*}$
can be measured in a theoretically clean way in \CP-violating $B$
meson decays to open-charm final states, $D^{(*)}X_s$. In these decays
the interference between the tree amplitudes $b\to c\bar u s$ and
$b\to u\bar c s$ leads to observables that depend
on the relative weak phase $\gamma$, the magnitude ratio $r_B\equiv\left|
\frac{A(b\to u)}{A(b\to c)}\right|$ and the relative strong phase
$\delta_B$ between the two amplitudes.
$r_B$ and $\delta_B$ depend on the $B$ decay under investigation; they
are difficult to estimate precisely from theory (QCD), but can be
extracted directly from data together with $\gamma$.

In order to search for the interference between the $b\to c\bar u s$
and $b\to u\bar c s$ tree amplitudes, we have reconstructed the
following decays, which lead to final states that can be produced both
through $b\to c$ and $b\to u$ mediated processes: (i) charged $B\to
D^{(*)0}/\overline{D}^{(*)0}K$ and $B\to D^0/\overline{D}^0 K^*
(K^*\to K^0_S\pi)$
decays and neutral $B^0\to D^0/\overline{D}^0 K^{*0} (K^{*0}\to
K^+\pi^-)$ decays, followed by $D^{(*)0}$ meson decays to final states
accessible also to $\overline{D}^{(*)0}$; (ii) neutral
$B^0/\overline{B}^0$ decays to $D^\pm K^0\pi^\mp$. In the
latter case, interference between the $b{\to}c$ and the $b{\to}u$
mediated processes is achieved through $B^0{-}\overline{B}^0$ mixing.

The measurements presented here exploit, partially or in total, the
large data sample ($465\times 10^6$
$B\overline{B}$ pairs, assumed to be equally divided into
$B^+B^-$ and $B^0\overline{B}^0$) accumulated by \babar\ in the years
1999-2007. Yet, they are still statistically limited, as the effects that
are being searched for are tiny:
\begin{itemize}
\item
The branching fractions of the $B$ meson decays considered here
are on the order of $5\times 10^{-4}$ ($B\to D^{(*)0}K$, $D^0K^*$ and
$DK^0\pi$) or $5\times 10^{-5}$ ($B^0\to \overline{D}^0K^{*0}$).
Branching fractions for $D^{(*)0}$ decays, including secondary decays,
range between $O(10^{-2})$ and $O(10^{-4})$; $D^-$ candidates are
reconstructed in $K^-\pi^+\pi^+$ ($\mathcal{B}\approx 10\%$).
\item
The interference between the $b\to c$ and $b\to u$
mediated $B$ decay amplitudes is predicted to be low, as the ratio
$r_B$ is expected to be around 40\% for neutral $B^0\to D^0K^{*0}$ and
$DK^0\pi$ (due to CKM factors) and around 10\% for charged $B\to
D^{(*)0}K^{(*)}$ decays (due to CKM factors and the additional
color-suppression of $A(b\to u)$).
\end{itemize}
The final states of these $B$ meson decays are completely reconstructed,
with efficiencies that range between 40\% (for low-multiplicity,
low-background decay modes) and 5\% (for high-multiplicity decays).
The selection is optimized in order to maximise the statistical sensitivity
$S/\sqrt{S+B}$, where the number of expected signal ($S$) and background
($B$) events is estimated from simulated samples and data control samples.

\section{MEASUREMENTS OF $\gamma$ WITH CHARGED $B$ MESON DECAYS}
\subsection{Dalitz-plot method\label{sec:dalitz_charged}}
On a sample of $383{\times}10^6$ $B\overline{B}$ pairs we have selected
$B{\to}D^0K$, $D^{*0}K$ ($D^{^*0}{\to}D^0\gamma$ and $D^0\pi^0$), and
$D^0K^* (K^*{\to}K^0_S\pi)$ decays, followed by $D^0$ decays to the
3-body self-conjugate final states $K^0_Sh^+h^-$
($h=\pi,K$).~\cite{bib:babar_DALITZ}
We have reconstructed 163 $B$ candidates with $D^0\to K^0_SKK$ and
979 $B$ candidates with $D^0\to K^0_S\pi\pi$.
The channel $B\to D^0K^*, D^0\to K^0_S K^+K^-$ has not been reconstructed
due to lack of statistics.
Following the approach proposed in \cite{bib:GGSZ},
from a fit to the Dalitz-plot distribution of
the $D^0$ daughters we have determined 2D confidence regions for the
variables $(x_\pm,y_\pm)$, which are related to $\gamma$ by
$x_\pm\equiv{r_B\cos(\delta_B\pm\gamma)}$ and
$y_\pm\equiv{r_B\sin(\delta_B\pm\gamma)}$.
In the fit we have used a model for the $D^0$ and $\overline{D}^0$
decay amplitudes to $K^0_Sh^+h^-$ described as the coherent sum of
a non-resonant part and several intermediate two-body decays that
proceed through known $K^0_Sh$ or $h^+h^-$ resonances. The
model has been determined from large ($\approx 5.6\times 10^5$) and very
pure ($\approx 98\%$) control samples of $D$ mesons produced in
$D^*{\to}D\pi$ decays.

\begin{figure}[!h]
\includegraphics[width=47mm]{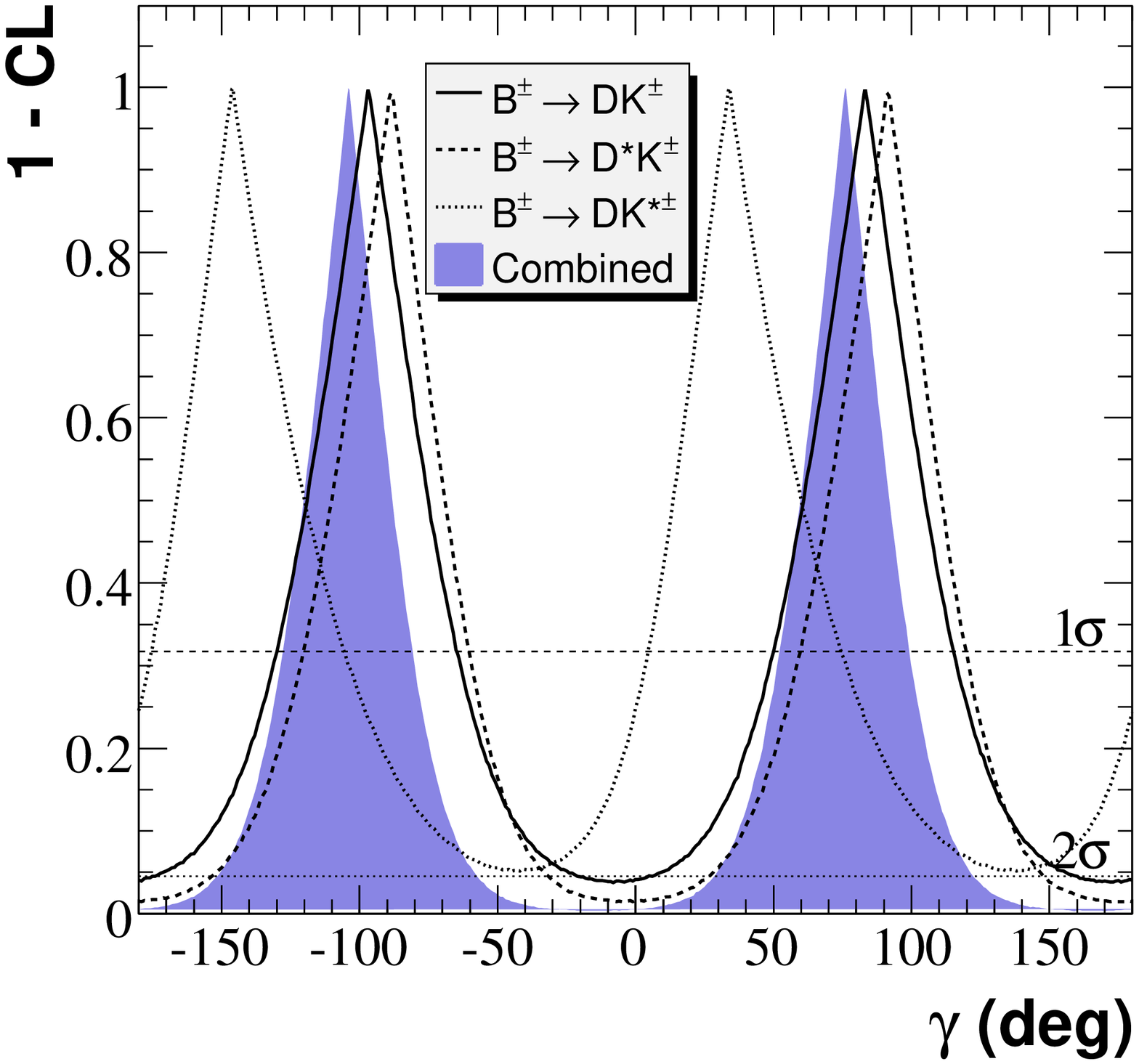}\includegraphics[width=47mm]{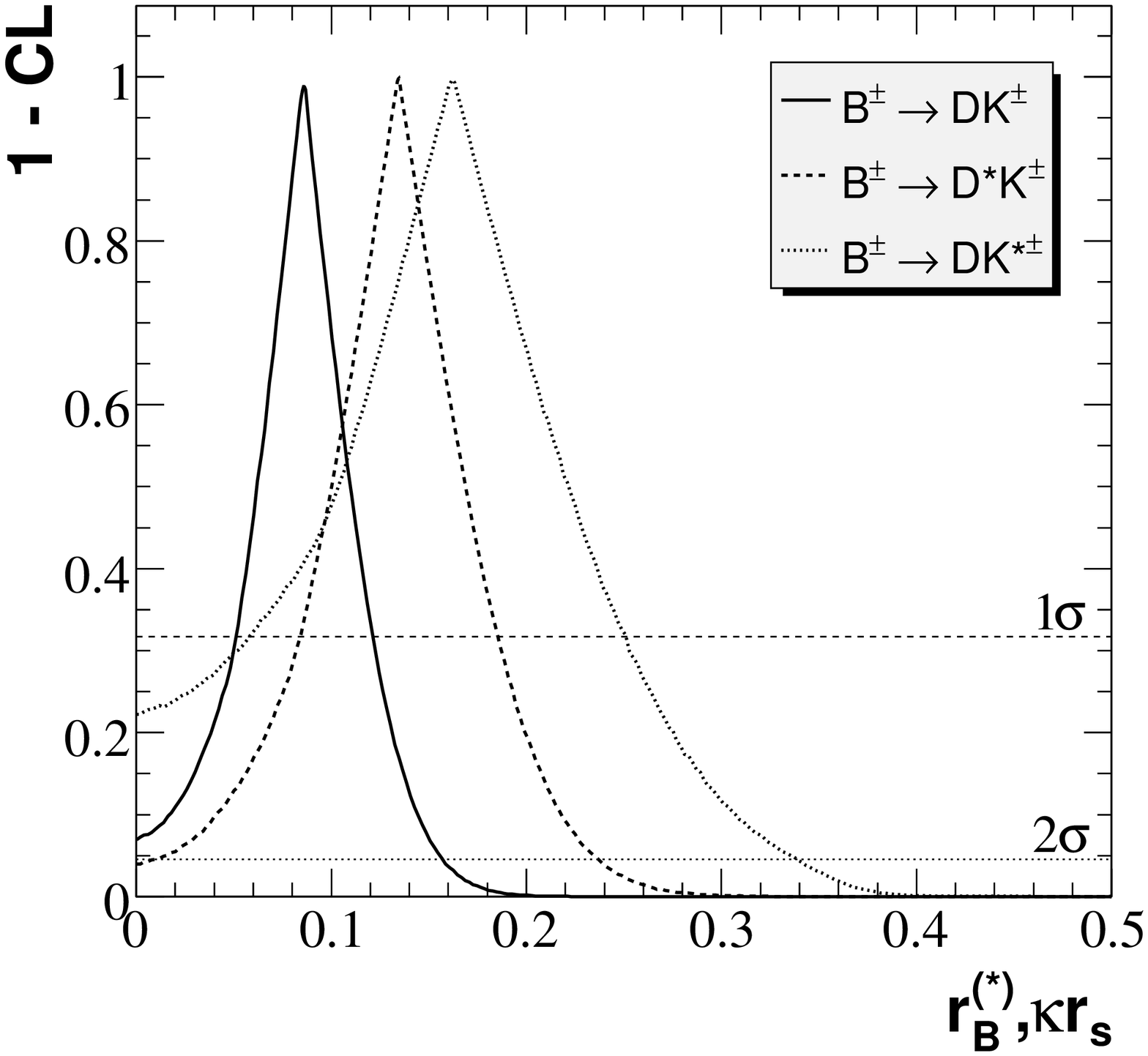}\includegraphics[height=53mm]{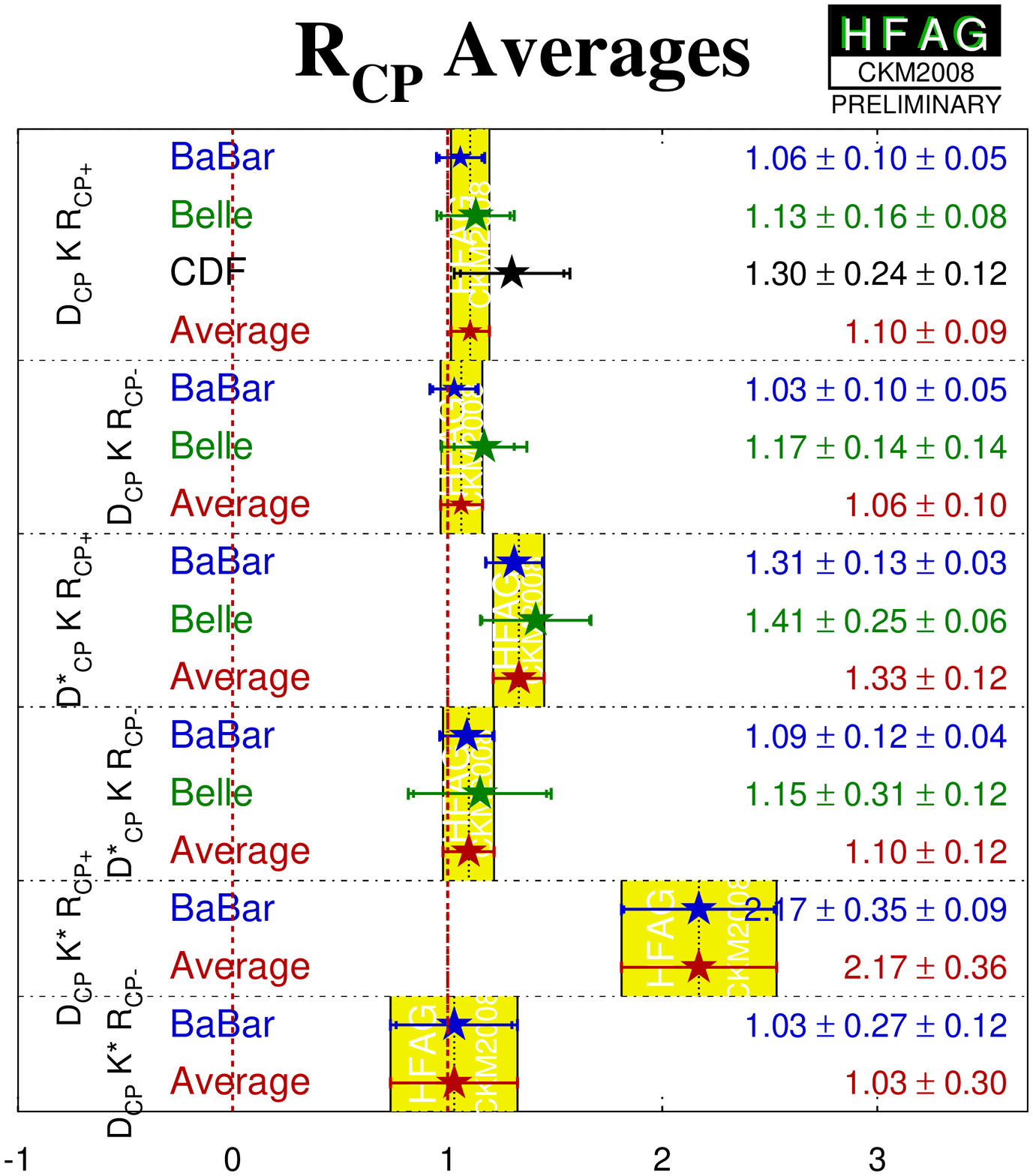}\includegraphics[height=53mm]{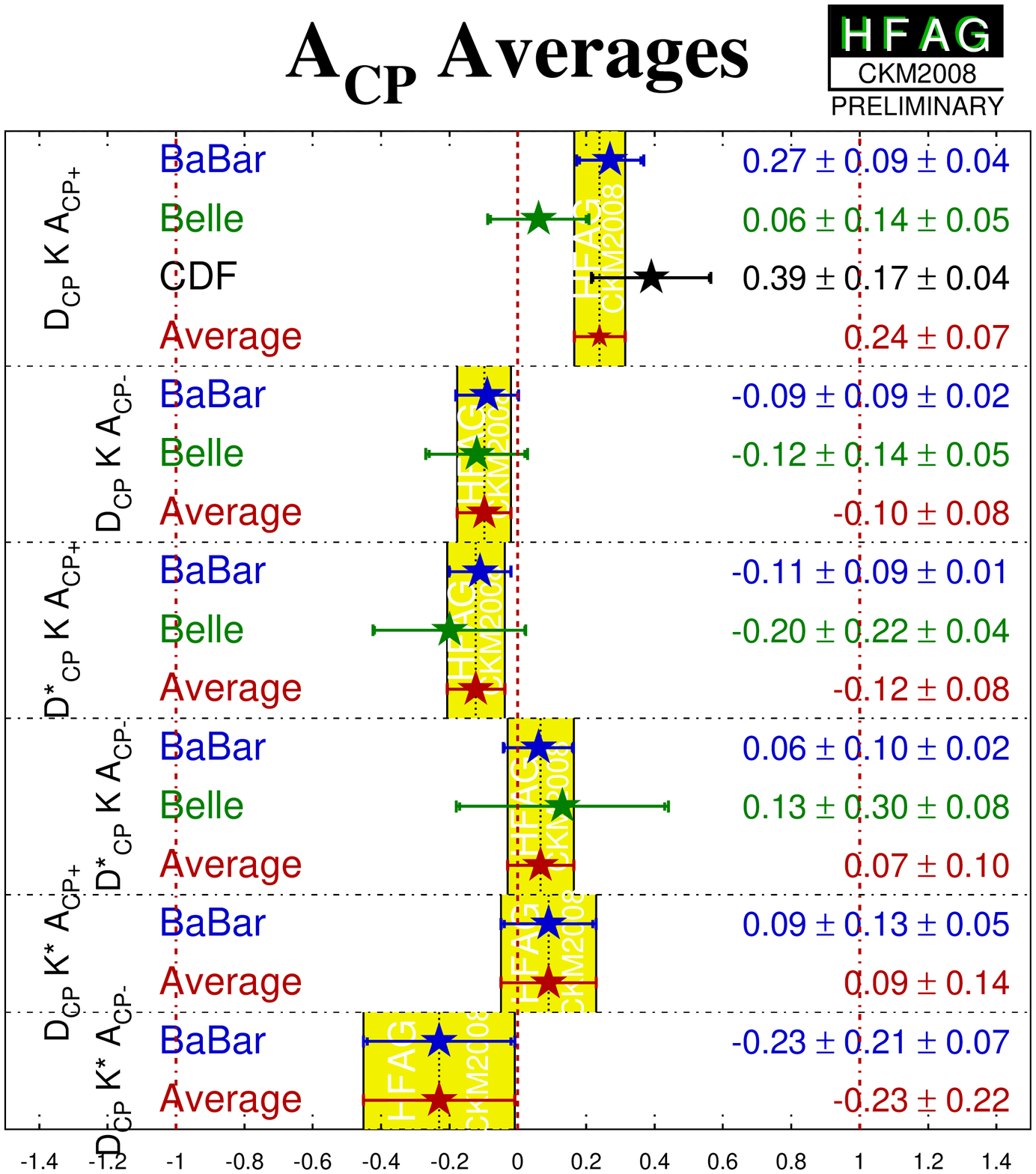}
\caption{Two left-most plots: 1-CL as a function of $\gamma$ and
    $r_B$ for $B\to D^0K$, $D^{*0}K$ and $D^0K^*$ as obtained with the
    Dalitz analysis of $D^0\to K^0_Sh^+h^-$. Two right-most plots:
    GLW observables measured by \babar\ and other
    experiments.\label{fig:charged_dk}}
\end{figure}

The results for $x$ and $y$ are summarized in Table~\ref{tab:xy}.
They are consistent with and have similar precision to those obtained
by the Belle experiment.~\cite{bib:belle_DALITZ}
\begin{table}[!h]
\begin{center}
\caption{Measurements of $x_\pm$ and $y_\pm$ obtained with the Dalitz
  and GLW analyses of $B\to D^{(*)0}K^{(*)}$}
\begin{tabular}{|l|c|c|c|}
\hline
\textbf{parameter} & $B\to D^0K$ & $B\to D^{*0}K$ & $B\to D^0K^*$ \\
\hline
$x_{+}$ (Dalitz) &
$-0.067\pm 0.043\pm 0.014\pm 0.011$ &
$\phantom{-}0.137\pm 0.068\pm 0.014\pm 0.005$ &
$-0.113\pm 0.107\pm 0.028\pm 0.018$ \\
$x_{+}$ (GLW) &
$-0.09\phantom{0}\pm 0.05\phantom{0} \pm 0.02\phantom{0} \phantom{ \pm
} \phantom{0..000}$ &
$\phantom{-}0.09\phantom{0} \pm0.07\phantom{0} \pm 0.02\phantom{0}
\phantom{\pm 0..000}$ & -  \\
$y_{+}$ (Dalitz) &
$-0.015\pm 0.055\pm 0.006\pm 0.008$ &
$\phantom{-}0.080\pm 0.102\pm 0.010\pm 0.012$ &
$\phantom{-}0.125\pm 0.139\pm 0.051\pm 0.010$ \\
$x_{-}$ (Dalitz) &
$\phantom{-}0.090\pm 0.043\pm 0.015\pm 0.011$ &
$-0.111\pm 0.069\pm 0.014\pm 0.004$ &
$\phantom{-}0.115\pm 0.138\pm 0.039\pm 0.014$ \\
$x_{-}$ (GLW) &
$\phantom{-}0.10\phantom{0}\pm 0.05\phantom{0}\pm0.03\phantom{0}
\phantom {+ 0..000}$ &
$-0.02\phantom{0} \pm0.06\phantom{0} \pm 0.02\phantom{0} \phantom{\pm
0}\phantom{.}\phantom{000}$ & - \\
$y_{-}$ (Dalitz) &
$\phantom{-}0.053\pm 0.056\pm 0.007\pm 0.015$ &
$-0.051\pm 0.080\pm 0.009\pm 0.010$ &
$\phantom{-}0.226\pm 0.142\pm 0.058\pm 0.011$ \\
\hline
\end{tabular}
\label{tab:xy}
\end{center}
\end{table}
From the $(x_\pm,y_\pm)$ confidence regions we determine, using
a frequentist procedure, $1\sigma$ confidence intervals
for $\gamma$, $r_B$ and $\delta_B$ (Fig.~\ref{fig:charged_dk}).
We obtain $\gamma\ {\rm mod}\ 180^\circ= (76^{+23}_{-24})^\circ$.
The total error is dominated by the statistical contribution:
the experimental and Dalitz-model-related systematic uncertainties
amount to $5^\circ$ each.
We find values of $r_B$ around 0.1, confirming that
interference is low in these channels:
$r^{D^0K}_B = 0.086{\pm}0.035$; $r^{D^{*0}K}_B = 0.135{\pm}0.051$;
$kr^{D^0K^*}_B=0.163^{+0.088}_{-0.105}$ ($k{=}0.9{\pm}0.1$ takes into
account the $K^*$ finite width). The small values of $r_B$
favored by our data are responsible - since $\sigma_\gamma{\approx}
\sigma_{x,y}/r_B$ - for the larger uncertainty on
$\gamma$ when compared to the analogous Belle
measurement, which favours values of $r_B$ about twice higher than
ours.
We also measure the strong phases (modulo $180^\circ$): $\delta^{D^0K}_B{=}(109^{+28}_{-31})^\circ$;
$\delta^{D^{*0}K}_B{=}(-63^{+28}_{-30})^\circ$;
$\delta^{D^0K^*}_B{=}(104^{+43}_{-41})^\circ$.
A $3\sigma$ evidence of direct \CP\ violation is found
when comparing $(x_+,y_+)$ to $(x_-,y_-)$ (they are equal
in absence of CPV) in the three $B$ decay channels.

\subsection{Gronau-London-Wyler (``GLW'') method}
We have reconstructed, on a sample of $383{\times}10^6$ $B\overline{B}$
pairs, $B{\to}D^0K$ and $D^{*0}K$ ($D^{^*0}{\to}D^0\gamma$ and
$D^0\pi^0$) decays with neutral $D$ mesons decaying to
\CP-even ($K^+K^-$, $\pi^+\pi^-$) and \CP-odd ($K^0_S\pi^0$, $K^0_S\phi$ and
$K^0_S\omega$) eigenstates. We have also reconstructed $B{\to}D^{(*)0}K$,
 $D^0{\to}K^-\pi^+$ and $B{\to}D^{(*)0}\pi$ decays as normalization channels.
We have identified almost 500 $B{\to}D^0_{\CP}K$~\cite{bib:babar_GLW_d0k}
and 500 $B{\to}D^{*0}_{\CP}K$~\cite{bib:babar_GLW_dstar0k}
decays. From the observed $B^\pm$ yields we have determined the
``GLW'' observables~\cite{bib:GLW},
$R^{(*)}_{\CP\pm}{\equiv}\frac{\Gamma(B^-{\to}D^{(*)0}_{\CP\pm}K^-)+\Gamma(B^+{\to}D^{(*)0}_{\CP\pm}K^+)}{(\Gamma(B^-{\to}D^{(*)0}K^-)+\Gamma(B^+{\to}\overline{D}^{(*)0}K^+))/2}
=  1{+}{r^{(*)}_B}^2{\pm}2r^{(*)}_B\cos\gamma\cos\delta^{(*)}_B$ and 
$A^{(*)}_{\CP\pm}{\equiv}\frac{\Gamma(B^-{\to}D^{(*)0}_{\CP\pm}K^-)-\Gamma(B^+{\to}D^{(*)0}_{\CP\pm}K^+)}{\Gamma(B^-{\to}D^{(*)0}_{\CP\pm}K^-)+\Gamma(B^+{\to}D^{(*)0}_{\CP\pm}K^+)}=\pm
2r^{(*)}_B\sin\gamma\sin\delta^{(*)}_B/R^{(*)}_{\CP\pm}$. 
Here we use the notation $r_B^{(*)}{\equiv}r_B^{D^{(*)0}K}$
and $\delta_B^{(*)}{\equiv}\delta_B^{D^{(*)0}K}$.
The results, summarized in Table~\ref{tab:GLW} and
Fig.~\ref{fig:charged_dk}, constitute the most precise measurements
of the $\CP$ asymmetries and branching fraction ratios in these decay
channels.
A $2.8\sigma$ hint of direct CPV is seen in $B\to D^0_{\CP+}K$ decays,
consistent with the results from the $B{\to}D^0K$,
$D^0{\to}K^0_Sh^+h^-$ Dalitz analysis and with recent results of the
GLW analysis of $B{\to}D^0_{\CP+}K$ decays performed by CDF.
From the relation
${r_B^{(*)}}^2{=}\frac{R^{(*)}_{\CP+}+R^{(*)}_{\CP-}-2}{2}$ we have
  obtained loose determinations of $r_B^{(*)}$:
$(r^{D^0K}_B)^2 = 0.05{\pm}0.07{\pm}0.03$, $(r^{D^{*0}K}_B)^2 = 0.22{\pm}0.09
 {\pm}0.03$.
We have also determined the variables $x_\pm$ by exploiting
the relations
$x_\pm{=}\frac{R_{\CP+}(1{\mp}A_{\CP+})-R_{\CP-}(1{\mp}A_{\CP-})}{4}$:
the results, consistent with and similar in accuracy to those from the
Dalitz analysis, are summarized in Table~\ref{tab:xy}.
Due to a discrete 8-fold ambiguity on $\gamma$ and the large
uncertainty on $r_B^{(*)}$, 
the GLW results alone do not constrain $\gamma$, but are
expected to improve the accuracy $\sigma_\gamma$ of the Dalitz results
presented in the previous section by a few degrees when combined with them.
\begin{table}[!h]
\begin{center}
\caption{Results of the $B\to D^0K$ and $B\to D^{*0}K$ GLW analyses}
\begin{tabular}{|l|c|c|c|c|}
\hline
\textbf{$B$ decay} & \textbf{$A^{(*)}_{\CP+}$} & \textbf{$R^{(*)}_{\CP+}$} &
\textbf{$A^{(*)}_{\CP-}$} & \textbf{$R^{(*)}_{\CP-}$} \\
\hline
$D^0K$ & $\phantom{-}0.27 \pm 0.09 \pm 0.04$ & $1.06\pm 0.10
\pm 0.05$ & $-0.09 \pm 0.09 \pm 0.02$ & $1.03\pm 0.10 \pm
0.05$\\
$D^{*0}K$ & $-0.11 \pm 0.09 \pm 0.01$ & $1.31\pm 0.13 \pm 0.04$ &
$\phantom{-}0.06 \pm 0.10 \pm 0.02$ & $1.10\pm 0.12 \pm 0.04$\\
\hline
\end{tabular}
\label{tab:GLW}
\end{center}
\end{table}

\section{MEASUREMENTS OF $\gamma$ WITH NEUTRAL $B$ MESON DECAYS}
\subsection{$B^0 \to \dbarp K^{*0}$: Atwood-Dunietz-Soni (``ADS'') and Dalitz-plot methods}
%
In the ``ADS'' analysis~\cite{bib:babar_ADS_d0kstar0} we
reconstruct $D^0$ decays to doubly-Cabibbo-suppressed final
states: $K^+\pi^-$, $K^+\pi^-\pi^0$ and $K^+\pi^-\pi^-\pi^+$. The
Cabibbo-allowed charge-conjugate final states are used as
normalization and control sample.
In principle $\gamma$ could be determined from the branching
fraction ratios ($R_{ADS}$) and the charge asymmetries ($A_{ADS}$) of
these decays.~\cite{bib:ADS}
In practice we do not have enough statistics to measure $A_{ADS}$
and put useful constraints on $\gamma$; on the other hand by measuring
$R_{ADS}$ we can infer a significant constraint on
$r_S{\equiv}r^{D^0K^{*0}}_B$.
We have selected, on a sample of $465\times 10^6$ $B\overline{B}$ pairs,
24 signal candidates summed over the 3 $D^0$ decay channels, and put 95\%
probability bayesian limits on $R_{ADS}$: $R_{ADS}^{K\pi}<0.244$;
$R_{ADS}^{K\pi\pi^0}<0.181$; $R_{ADS}^{K\pi\pi\pi}<0.391$.
We have then
combined those 3 results and determined bayesian probability
regions for $r_S$ by adding external information (from $c$- and
$B$-factories) on the $\dbarp$ decay 
amplitudes. At 95\% probability $r_S$ lies between 7 and
41\% (Fig.~\ref{fig:neutral_dkpi}).

In the Dalitz analysis~\cite{bib:babar_DALITZ_d0kstar0} we
reconstruct $\dbarp$ decays to $K^0_S\pi^+\pi^-$.
Using $371 \times 10^6$ $B\overline{B}$ pairs we
have selected 39 signal candidates. From the fit to the Dalitz-plot
distribution of the $D^0$ daughters, using the same model for the
$\dbarp{\to}K^0_S\pi^+\pi^-$ decay amplitudes as in
Sec.~\ref{sec:dalitz_charged}, we have
determined 2D bayesian probability regions for $(r_S,\gamma)$
(Fig.~\ref{fig:neutral_dkpi}). After combination
with the $r_S$ likelihood from the ADS method we obtain, at
68\% probability, $\gamma\,{\rm mod}\,180^\circ{=}(162{\pm}56)^\circ$.

\subsection{$2\beta+\gamma$ from a time-dependent Dalitz-plot analysis
  of $B^0/\overline{B}^0 \to D^\pm K^{0}\pi^\mp$}
On a sample of $347\times 10^6$ $B\overline{B}$ pairs we have searched
for neutral $B$ meson decays to $D^\pm
K^0_S\pi^\mp$.~\cite{bib:babar_DALITZ_dkpi}
We retain only those events where the other $B$ meson ($B_{\rm tag}$)
decays to a final state which allows to determine its flavor
(either $B^0$ or $\overline{B}^0$) and the proper time difference
$\Delta t$ between the two $B$ decays, inferred from the separation of
the $B$ decay vertices and momenta.
The effective efficiency of the flavor-tagging algorithm is around $30\%$.
A fit to the Dalitz-plot distribution of $DK^0_s\pi$ as a function of
$\Delta t$ yields $2\beta{+}\gamma$ through the relations:
$\Gamma(\vec{x},\Delta t, \xi,
\eta)=\frac{A_c(\vec{x})^2+A_u(\vec{x})^2}{2}\times
\frac{e^{-\frac{|\Delta t|}{\tau_B}}}{4\tau_B} \times \{1-\eta\xi
C(\vec{x})\cos(\Delta m_d \Delta t)\ + \xi S_\eta(\vec{x})\sin(\Delta
m_d \Delta t)\}$, $S_\eta(\vec{x})=\frac{2\Im(A_c(\vec{x})A_u(\vec{x})e^{i(2\beta+\gamma)+\eta i (\phi_c(\vec{x})-\phi_u(\vec{x}))})}{A_c(\vec{x})^2+A_u(\vec{x})^2}$,
$C(\vec{x}){=}\frac{A_c(\vec{x})^2-A_u(\vec{x})^2}{A_c(\vec{x})^2+A_u(\vec{x})^2}$.
$\tau_B$ is the $B^0$ lifetime, $\Delta m$ the $B^0-\overline{B}^0$
mixing frequency, $\vec{x}$ the position in the $DK\pi$ Dalitz plot;
$\xi{=}-1(1)$ if the flavor of the $B_{\rm tag}$ is $B^0 (\overline{B}^0)$,
and $\eta{=}+1(-1)$ if the final state contains a $D^+ (D^-)$.

\begin{figure}
\includegraphics[height=40mm,width=52mm]{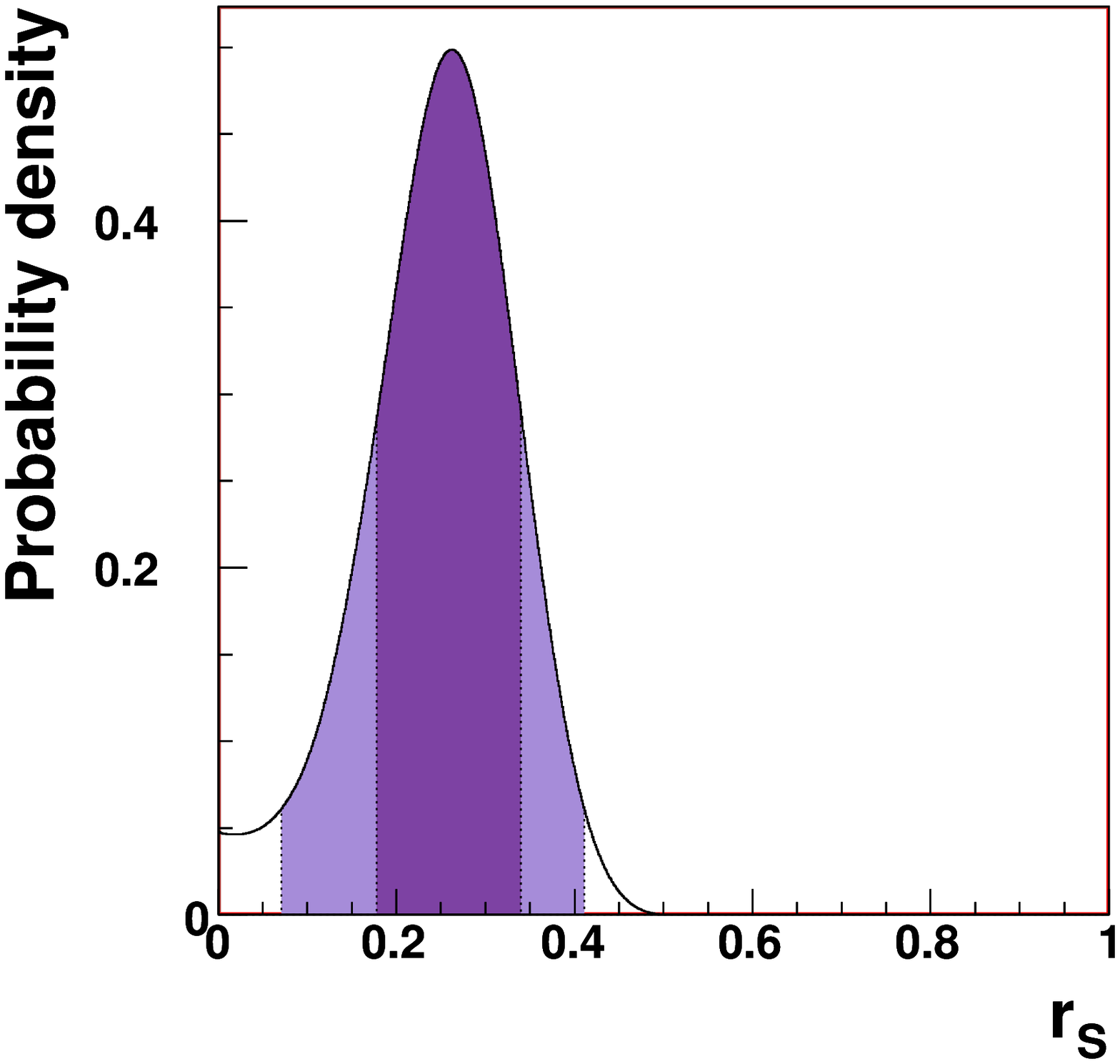}\includegraphics[height=40mm,width=68mm]{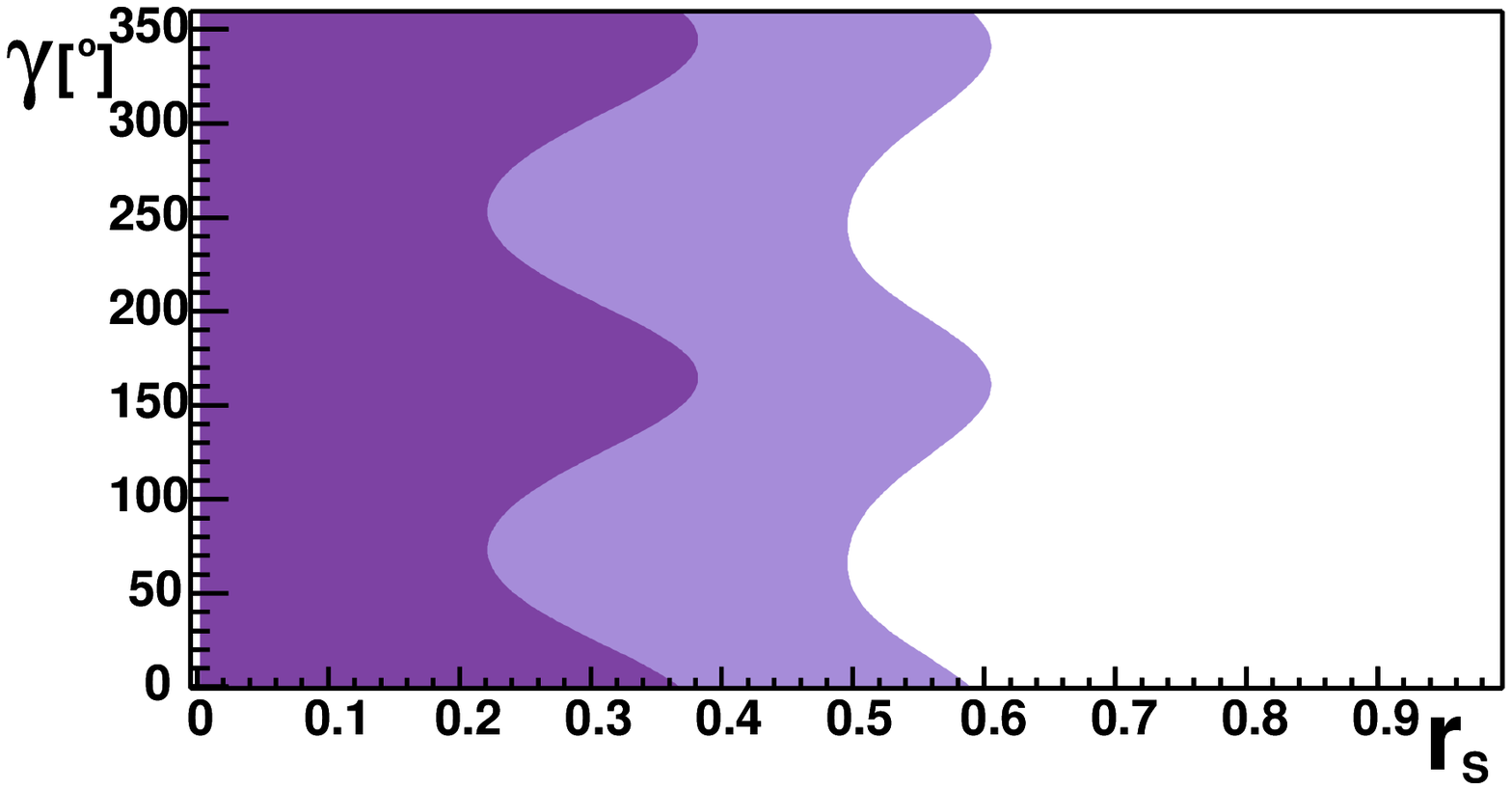}\includegraphics[height=40mm,width=53mm]{scan_r.epsi}
\caption{\label{fig:neutral_dkpi}Left: likelihood function for
  $r_{S}$ from the ADS analysis of $B^0{\to}D^0K^{*0}$, including
  68$\%$ (dark-shaded) and 95$\%$ (light-shaded) probability regions. 
  Center: $\gamma$ vs $r_S$ 68\% and 95\% probability regions obtained from the
  $B^0{\to}D^0K^{*0}$ Dalitz analysis. Right: distribution of the
  fitted values of $2\beta + \gamma$ from selected $B^0\to DK^0\pi$
  decays for different hypotheses on the value of $r$.}
\end{figure}

We have selected 558 signal candidates and performed the
time-dependent Dalitz-plot fit.
For the direct $B$ decay amplitude we use a model which incorporates all known
intermediate resonances and float most of the magnitudes ($A$) and
phases ($\phi$) in the fit.
The ratio $r$ between the $b{\to}u$ ($A_u$) and $b{\to}c$ ($A_c$)
amplitudes is fixed.
A scan of $2\beta{+}\gamma$ as a function of $r$ is shown in
Fig.~\ref{fig:neutral_dkpi}.
For the final $2\beta + \gamma$ result we assume $r=0.3\pm0.1$.
Using the world average for $\beta$ from $B$ decays to
charmonium we obtain $\gamma\ {\rm mod}\ 180^\circ=(40\pm
57)^\circ$, in agreement with the other determinations.

\section{CONCLUSION}
Many new results on $\gamma$ have been obtained by \babar\ during
2008.
They are in good agreement with each other and favor
$\gamma{\approx}72^\circ$, dominated by the
$B{\to}D^{(*)0}K^{*}$, $D^0{\to}K^0_Sh^+h^-$ Dalitz analysis and
consistent with expectations from CKM fits.
Thanks to the combination of several strategies, the uncertainty
$\sigma_\gamma$ is approaching $20^\circ$, which was not an original
goal of the $B$ factories.
$\sigma_\gamma$ is limited by the available statistics; we expect
a 20\% improvement from the collected data that have not been analyzed
yet.
The interference effects are confirmed to be small both
in charged and neutral $B$ decays, thus a much larger statistics is
needed to reach a precision of a few degrees on $\gamma$.



\end{document}